\documentstyle[multicol, prb,aps,epsfig]{revtex}
\newcommand{\rb}{\mbox{\boldmath$r$}}
\newcommand{\pb}{\mbox{\boldmath$p$}}
\newcommand{\om}{\mbox{\boldmath$\Omega$}}
\begin{document}
\date{\today}
\title{Criticality in polar fluids}
\author{\bf Yan Levin$^{1,2}$\footnote{Corresponding author}, 
Paulo S. Kuhn$^1$, Marcia C. Barbosa$^1$}
\address{\it $^1$Instituto de F\'{\i}sica, Universidade Federal
do Rio Grande do Sul\\ Caixa Postal 15051, CEP 91501-970, 
Porto Alegre, RS, Brazil\\ 
$^2$Institute for Theoretical Physics, University of California,\\
Santa Barbara, CA 93106-2431, USA\\
{\small levin@if.ufrgs.br}}
\maketitle
\begin{abstract}
A model of polar fluid is studied theoretically. The interaction potential,
 in addition to dipole-dipole term, possesses a 
dispersion contribution of the
van der Waals-London form.  It is found that when the dispersion force
is comparable to  dipole-dipole interaction, the 
fluid separates into  coexisting
liquid and gas phases. The calculated critical parameters are in excellent
agreement with Monte Carlo simulations.  When 
the strength of dispersion attraction is bellow critical, 
no phase separation is found. 

\end{abstract}

\centerline{{\bf PACS numbers:} 61.20.Gy; 64.70.Fx; 75.50.Mm;}
\bigskip


The phase behavior of simple molecular fluids was thought
to be well understood since the early work of van der Waals
over a century ago.  It came, therefore, as  a  shock  when
the simulations in the early $90's$ could not locate
the anticipated liquid-gas coexistence in the simplest model
of polar fluid, idealized by dipolar hard spheres $DHS$
 \cite{Wei93,Lee93,Ste94}. 
The shock was further augmented
by the fact that our everyday experience with polar fluids, such as water,
does not leave any doubt that they do posses both liquid and
gas phases.  Furthermore, the theoretical methods such as, 
integral equations \cite{Wer71}
and thermodynamic perturbation theory\cite{Gen70,Rus73}, 
developed over the hundred years
since van der Waals first proposed his celebrated theory, all predicted
such a phase separation.  What could have  gone wrong?

The simulations of particles which interact by long ranged potentials,
such as Coulomb or dipolar, are notorious for their difficulty.  Thus,
it is always possible that for some technical reason the simulations could not 
access the region of instability.  
This, however, seems to be less and 
less likely in view of constantly 
improving computational power and algorithm design.  
The first hint of what was happening was provided by the original
simulations of Weis, Levesque, and Caillol\cite{Wei93}. 
These authors observed that at low temperatures,
where  the theories predicted phase separation  
into coexisting liquid and gas phases,
the dipolar fluid became highly structured, with particles
forming chains of aligned dipoles. 
A further clue was obtained by van Leeuwen and Smit\cite{Lee93} who simulated
soft dipolar particles in which the softcore repulsion, 
$1/r^{12}$, was augmented by 
a variable isotropic attraction of  
 van der Waals-London form\cite{Gro96}, $1/r^6$.  
By performing the Gibbs-ensemble
Monte Carlo simulation van Leeuven and Smit came to the conclusion
that a minimum strength of isotropic attraction was necessary
in order to produce phase separation. The ubiquity of
liquid-gas coexistence in real world can, therefore, be attributed to the
presence of dispersion interactions between the
molecules of polar fluids. A number of
theories have been put forward to try to explain the unusual behavior
found in computer simulations\cite{Sea96,Roi96,Tav97}, 
but none has been able to completely 
resolve the mystery.  
Recently a new theory based on the
pioneering ideas of  Debye, H\"uckel\cite{Deb23}, Bjerrum\cite{Bje26}, 
and Onsager\cite{Ons36}
 has been advanced to account for the absence 
of phase separation in  $DHS$ \cite{Lev99}. 
A similar approach has proven to be successful in the study of coulombic
criticality in restricted primitive model ($RPM$) of 
electrolyte solutions\cite{Fis93},
predicting a coexistence curve in good agreement with experiments
and computer simulations\cite{Val91}.
From the theoretical perspective, however, the 
system simulated by van Leeuven and Smit\cite{Lee93} is
significantly more complex than the $RPM$ or the $DHS$.  Besides
the dipole-dipole interaction the theory must be able to deal with the 
softcore repulsion, as well as with the short ranged dispersion
attraction.  This model, therefore, provides a very stringent
test of any theory of criticality in polar fluids.  
A successful theory
should account for both, the presence of phase 
separation for strong dispersion forces, 
as well as for its disappearance when the isotropic 
coupling is reduced bellow the critical value.

Our model, then, consists of $N$ molecules of dipole moment $\pb$,
inside a uniform medium of dielectric constant $\epsilon_0$.
The interaction potential between any two particles is

\begin{equation}
\label{1}
U(\rb;\pb_1,\pb_2)=\frac{p^2}{\epsilon_0 \sigma^3}
\left(\left(\frac{\sigma}{r}\right)^{12}
-\lambda \left(\frac{\sigma}{r}\right)^6\right)+
\frac{1}{\epsilon_0 r^3}\left(\pb_1\cdot\pb_2-
\frac{3(\pb_1\cdot\rb)(\pb_2\cdot\rb)}{r^2} \right)\;.
\end{equation}
The dimensionless 
parameter $\lambda$ defines
the strength of the isotropic interaction as it compares with
dipole-dipole electrostatic potential. To proceed we associate with the
 model of 
``soft'' molecules given by Eq.~(\ref{1}) an equivalent system of
``rigid'' particles. The advantage of working with hard particles
is that a number of analytic  equations of state have been
developed to study rigid molecules, in particular hard spheres. Fixing one
molecule at the origin with the dipole moment aligned with the
z-axis, and a second particle  at $\rb$ with the 
dipolar orientation $\om$,  the
potential in Eq.~(\ref{1}) can be separated in two parts\cite{Wee71}. 
The repulsive term, $U_{rp}(\rb,\om)
=U(\rb;\pb_1,\pb_2)-U(\rb_{min},\om_{min})$ for 
$\vert\rb\vert<\vert\rb_{min}\vert$ and $U_{rp}(\rb,\om)=0$ for 
$\vert\rb\vert \geq \vert\rb_{min}\vert$; and the attractive term,
$U_{at}(\rb,\om)=U(\rb_{min},\om_{min})$ for 
$\vert\rb\vert<\vert\rb_{min}\vert$ and 
$U_{at}(\rb,\om)=U(\rb;\pb_1,\pb_2 )$ for 
$\vert\rb\vert \geq \vert\rb_{min}\vert$.
Here $\rb_{min}$ is the intermolecular vector and
$\om_{min}$ is the dipolar orientation 
for which the interaction 
potential is minimum. In the spirit of Weeks, Chandler, and
Andersen (WCA)\cite{Wee71} we can associate with the soft
particles given by Eq.~(1) a system of rigid dipoles interacting by
\begin{eqnarray}
\label{2}
U^{dd}(\rb)&=&-\frac{\lambda p^2 a^3(\theta;\lambda)}{\epsilon_0 r^6 }+
\frac{1}{\epsilon_0 r^3}\left(\pb_1\cdot\pb_2-
\frac{3(\pb_1\cdot\rb)(\pb_2\cdot\rb)}{r^2} \right)\;
\rb \geq \vert {\bf a} \vert  \nonumber \\ 
U^{dd}(\rb)&=&\infty \;\; \rb < \vert {\bf a} \vert \;,
\end{eqnarray}
where the distance of 
closest approach $a(\theta;\lambda)$ is determined by the
condition, $U_{rp}({\bf a},\om_{min})=k_BT$. In view of the azimuthal
symmetry, $\vert {\bf a} \vert=a(\theta;\lambda) $ is only a function of
the angle $\theta$ between the dipole moment $\pb_1$ and the
intermolecular vector $\rb$.  To stress that this distance
depends on the strength of the isotropic attraction, we have
included $\lambda$ as an explicit parameter in $a(\theta;\lambda)$.
 The excluded volume
region around the central dipole has the form of geoid, a flattened sphere,  
with
the eccentricity a decreasing function of $\lambda$ (the highest eccentricity
for $\lambda=0$).
 
The reduced free energy density, $f=\beta F/V$,  
of a fluid with the interaction potential Eq.~\ref{2} can be constructed as a
sum of terms embodying the most relevant physical features, starting
with the entropic ideal gas contribution, 
$f^{id}=\rho \ln(\rho \Lambda^3)-\rho$.  Here $\rho=N/V$ is the density
of dipoles, $\beta=1/k_BT$, and  $\Lambda$ is the thermal wavelength. 
The mean energy of interaction
between the dipoles can be calculated as
\begin{equation}
\label{3}
F^{vdW}=\frac{1}{2}\int d{\bf r_1} d{\bf r_2}
\rho({\bf r_1})\rho({\bf r_2})U^{dd}({\bf r_1},{\bf r_2})\,,
\end{equation}
where $\rho({\bf r})$ is the local density of particles and where,
to simplify the notation, we have suppressed integration over the relative
orientation of dipole moments.  
Since the distribution of particles and their orientation
is uniform, only the first term of Eq.~(\ref{2})
contributes to the integral, with the second term averaging
to zero. In order to simplify the integration we can replace the
excluded volume region by a sphere with radius 
$a_s(\lambda)=(a(0;\lambda)+2 a(\pi/2;\lambda))/3$. 
Since the eccentricity is quite small 
this is a good approximation. We have numerically checked
that the error introduced by this approximation is no more
than a few percent. Performing the integration, we find the reduced free
energy due to dispersion interactions to be
\begin{equation}
\label{4}
f^{vdW}=-\frac{2 \pi}{3a_s^3(\lambda)}\frac{\lambda (\rho^*)^2}{T^*} \;,
\end{equation}
where we have defined the reduced (dimensionless) temperature and density 
to be, $T^*=k_B T\epsilon_0 \sigma^3/p^2$ and $\rho^*=\rho \sigma^3$, 
respectively.

Although {\it on average} the dipoles are distributed uniformly, with
no particular preference for any specific orientation, the long ranged
dipolar force  produces strong correlations between the dipole moments of
individual particles.  This leads to a significant contribution
to the total free energy.   To obtain this correlational
free energy, we fix one
particle at the origin with its dipole moment aligned with the z-axis. 
The electrostatic potential that this dipole 
feels due to the presence of other molecules 
can be found from the solution of Laplace equation
$\nabla^2 \phi=0$. The boundary conditions require continuity
of the potential, $\phi_{in}(a_s)= \phi_{out}(a_s)$, 
and the displacement field, 
$\epsilon_0 \phi'_{in}(a_s)=\epsilon \phi'_{out}(a_s)$,
across the surface $r=a_s$.
The susceptibility of the outer 
dipoles to polarization by the electric field can 
be characterized by the renormalized dielectric constant $\epsilon$,
the expression for which can be obtained
from the Onsager's reaction field theory\cite{Ons36},
$(\epsilon-\epsilon_0)(\epsilon_0+2 \epsilon)/ \epsilon= 
4 \pi\beta p^2 \rho$.
 The Laplace equation can now be integrated
to yield the electric field felt by the central dipole due to 
the other particles,
\begin{equation}
\label{6}
{\bf E}_0=\frac{2 p}{\epsilon_0 a_s^3(\lambda)} 
\frac{\epsilon-\epsilon_0}{2 \epsilon+\epsilon_0}\hat{\bf z}\;.
\end{equation}
The electrostatic {\it free energy} of
the whole system is obtained from the Debye charging 
process\cite{Deb23,Fis93}
in which all the particles are charged simultaneously
from zero to their final dipolar strength.
The integration can be done explicitly
yielding the correlational free
energy density~\cite{Lev99,Nie71},
\begin{equation}
\label{7}
f^{dd}
= -\frac{1}{4\pi a_s^3(\lambda)} 
\left\{-2+\frac{1}{\psi(u)}+\psi(u)+\frac{9}{2}
\ln\left( \frac{3}{2\psi(u)+1}\right) +3 \ln\psi(u)\right\}
\end{equation}
with $\psi(u) \equiv \epsilon(u)/\epsilon_0 =\frac{1}{4}\left(1+u\right)
+\frac{1}{4}\sqrt{9+2u+u^2}$,
and $u=4\pi\rho^*/T^*$.  
Finally, the contribution to the total free energy due to hardcore
repulsion can be approximated by the Carnahan-Starling ($CS$) form\cite{Car69}, 
$f^{hc}=\rho g(\eta)$, where $g(\eta)=\eta (4-3\eta)/(1-\eta)^2$
and $\eta=\pi a_s^3(\lambda) \rho /6$ is the volume fraction occupied
by dipoles.

Combining all
of the contributions, the total free energy
of dipolar fluid becomes $f=f^{id}+f^{hc}+f^{vdW}+f^{dd}$. It is a simple
matter to see that for any $\lambda$ this free energy 
violates the convexity requirement when the temperature is lowered below the
critical value $T_c(\lambda)$.  The critical parameters are plotted
in Fig. 1.  It is evident that the agreement with the simulations is
quite good.  For $0.3<\lambda \leq 1$, the critical densities
are close to the ones obtained in the simulations.
\begin{figure}[h]
\epsfig{file=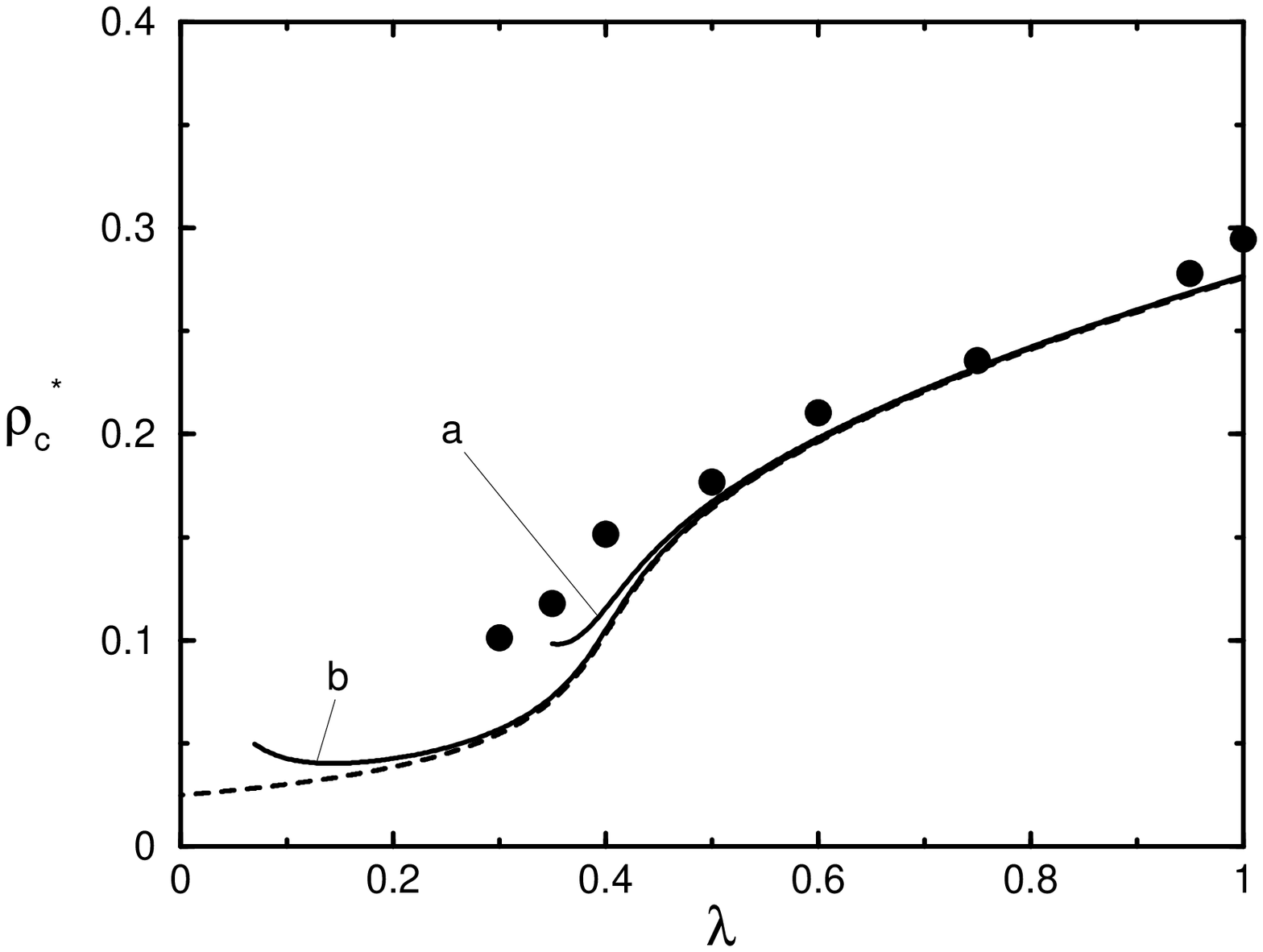,width=7.2cm}
\epsfig{file=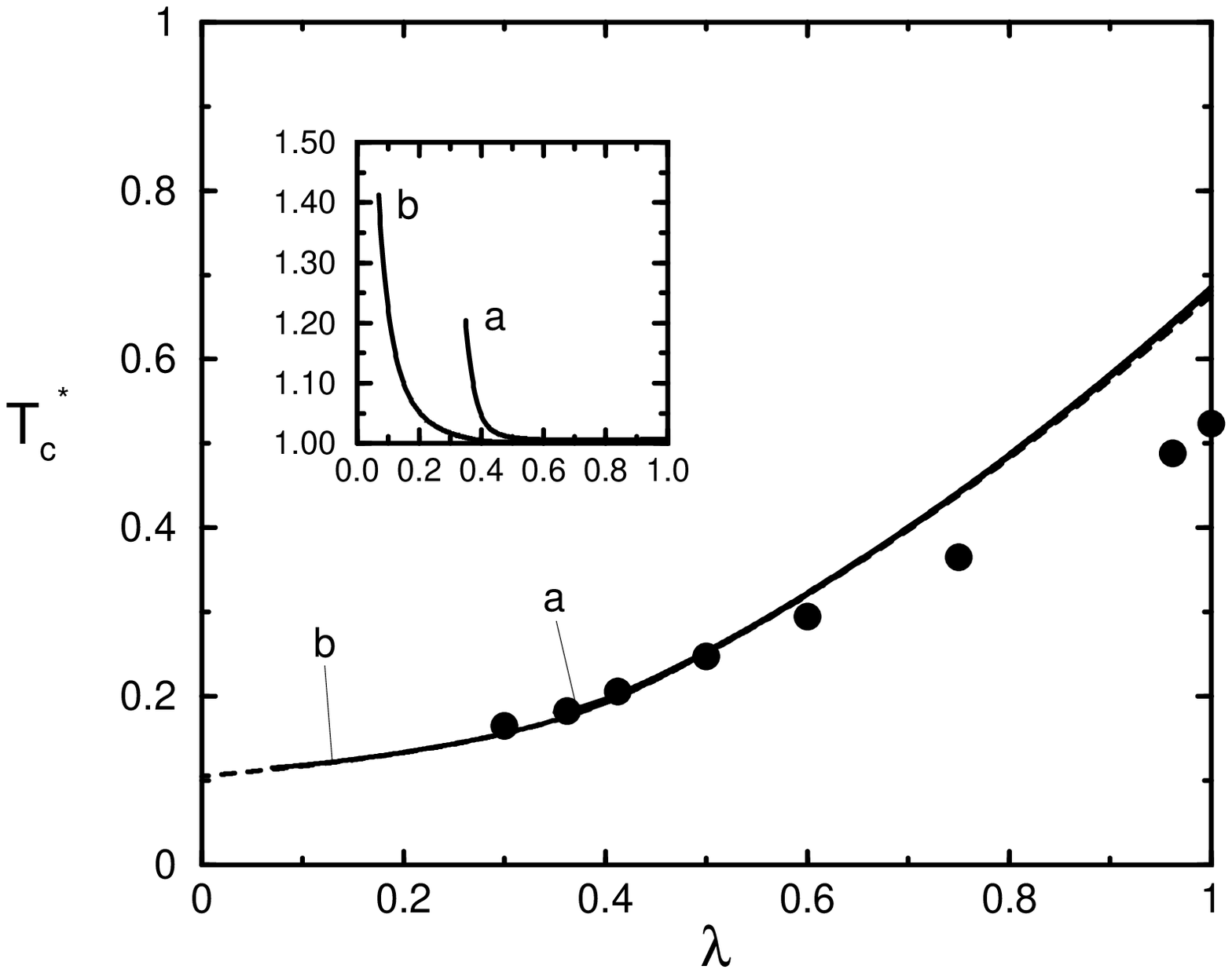,width=7.2cm}
\caption{Critical density
and critical temperature as a 
function of $\lambda$: the dashed curve is for the theory without
association; the two solid curves $a$ and $b$ are for the limiting cases 
$h(n) \approx 1.52165...$ and  $h(n) \approx 1.19055...$, 
respectively. Solid circles 
are the Monte Carlo data from Ref[\cite{Lee93}].  
Inset shows the average cluster size as a
function of $\lambda$.}
\label{Fig1}
\end{figure}
For $\lambda=1$ the critical temperature is
about $30\%$ too high.  However, for smaller isotropic 
couplings, $\lambda< 0.6$, 
this discrepancy almost completely disappears. It is not surprising that
for $\lambda \approx 1$ the agreement is the poorest.  After all, the theory
presented above is intrinsically mean-field, and as such neglects the thermal
fluctuations which are the strongest in the vicinity of the critical point.
For short ranged potentials,
the effect of fluctuations is to significantly depress the critical
temperature.  Even for such relatively simple system as ``argon''
mean-field theories overestimate $T_c$ by $15\%$ to
$20\%$\cite{Fis94}. We might expect that as the relative strength 
of isotropic short ranged 
interaction decreases,
compared to the long ranged dipolar force, the effect of fluctuations
on the critical temperature will also diminish.  This, indeed, is 
the case, see Fig. 1. We note that for $\lambda \approx 1$ 
the phase separation is the result of the
competition between the attractive and the excluded volume, $CS$, interactions.
For small $\lambda 's$, on the other hand, the instability is purely 
electrostatic and is only weakly affected by the hardcore contribution
to the total free energy.

For $\lambda<0.3$ the simulations fail to find any coexistence,
while our theory predicts phase separation all the way down to $\lambda=0$.
What is responsible for this discrepancy?  In view of the 
previous discussion, we do not have to look too far for
an explanation. It was observed in simulations 
that for $\lambda<0.3$, the low temperature thermodynamics was dominated
by  polymer-like chains of aligned dipoles.  On the other hand,
the theory presented above is intrinsically linear, and does not
take into account the strong non-linear effects associated
with the formation of clusters. To account for these, while preserving
the theory's linear structure, we shall explicitly
allow for chains composed of $n$ associated dipoles. Based on
the Monte Carlo simulations 
we shall ignore other possible geometries of clusters.  This, actually, is
a non-trivial approximation, since at zero temperature
the energy favors compact clusters\cite{Lev99}.  The simulations, however,
find that for finite temperatures linear chains predominate. 
Thus, as a first approximation we shall only consider polymer-like
clusters.

The  density of chains containing $n$ monomers will be denoted by $\rho_n$,
with the density of free, unassociated, dipoles $\rho_1$.
The particle conservation leads to
$\rho=\sum_{n=1}^{\infty} n \rho_n$,
where the total density of dipoles is  $\rho=N/V$.
This is the so called ``chemical'', as opposed to ``physical'' 
picture of statistical mechanics of strongly
interacting particles.  It is important to keep in
mind that no new approximations are being introduced into
the theory, but only a change of perspective.  Clearly if, as it happens
at high temperatures, the formation of clusters is unlikely, the theory
should inform us of this fact by predicting low cluster density.  
As was noted by Onsager ``what we remove from one page of
the ledger would be entered elsewhere with the same effect'' \cite{Fal71}.  
In fact it is possible to directly
map this chemical picture onto exact thermodynamic perturbation
theory\cite{Hil56}.  The new perspective is especially useful at low
temperatures, where the perturbation theory is very difficult to construct,
but where the ``chemical'' view point provides a particularly clear
perspective  of the physical phenomena.   We can now proceed to make the first
approximation. In the spirit of Bjerrum\cite{Bje26}, we shall assume that
the clusters interact only weakly with each other and with the
free dipoles.  Thus, as a second approximation, we shall treat
clusters as ideal non-interacting species with all of the
interactions confined to free dipoles. This
approximation has proven to work quite well in 
the theory of electrolytes, in particular allowing for a precise location
of the critical point\cite{Fis93}.

The entropic free energy due to thermal motion of dipoles  
and clusters can be approximated by 
the ideal gas contribution, 
$f^{id}=\sum_{n=1}^{\infty}\rho_n \ln(\rho_n \Lambda^{3n}/\zeta_n)-\rho_n$.  
In the limit of low temperatures, the internal partition 
function of an $n$ cluster, $\zeta_n$,
can be evaluated for chainlike configurations to 
yield\cite{Gen70,Lev99,Jor73}
\begin{equation}
\label{12}
\zeta_n=\left\{\frac{\pi^{3/2}2^{21/18} T^{*5/2} 
\sigma^3}{9 \sqrt{3}}\right\}^{n-1}
\exp\left\{\frac{(n-1) h(n)}{T^*}\right\} \;
\end{equation}
where,
\begin{equation}
\label{13}
h(n)=3 \left[\frac {n[\psi^{(2)}(n)-\psi^{(2)}(1)]
+2[\psi^{(1)}(n)-\psi^{(1)}(1)]}{4(n-1)}\right]^{4/3}\;.
\end{equation}
The $\psi^{(1)}(n)$ and $\psi^{(2)}(n)$ are the polygamma functions
of the first and second order, respectively.
For free dipoles $\zeta_1=1$. The hardcore contribution can, once
again, be approximated by the Carnahan-Starling form, 
$f^{hc}=\sum_{n=1}^\infty \rho_n g(\eta)$, where 
$\eta=\pi a_s^3(\lambda) \rho /6$ is the total volume fraction occupied
by dipoles and chains.
The mean-field and the electrostatic contributions to the
free energy are given by  Eqns.~(\ref{4}) and (\ref{7}), with the
total density $\rho$ replaced by the density of free dipoles $\rho_1$.
The distribution of cluster densities can be obtained through 
minimization of the total free energy $f=f^{id}+f^{hc}+f^{vdW}+f^{dd}$,
 subject to constraint of particle conservation.
 This reduces to the law of mass action,  $\mu_n=n \mu_1$, relating the
chemical potential of free dipoles $\mu_1=\partial f/\partial \rho_1$,
to the chemical potential of  n clusters $\mu_n=\partial f/\partial \rho_n$.
Substituting the free energy, the low of mass action yields
\begin{equation}
\label{14}
\rho_n=\zeta_n \rho_1^n e^{\beta n \mu^{ex}+(n-1)g(\eta)} \;,\; n \geq 2\;;
\end{equation}
where the excess chemical potential is 
$\mu^{ex}=\partial f^{dd}/\partial \rho_1$. 
In spite of its simple appearance the law of mass action is actually 
an infinite set of coupled algebraic equations which determine the
distribution of chain densities $\{\rho_n\}$.  To solve these
equations is a difficult numerical task. In order
to proceed we observe that  $h(n)$ is a uniformly increasing function
bounded by the two limiting values, 
$h(2)=1.19055...$ and $h(\infty)=1.52165...$;
{\it i.e.} $ 1.19055... \leq h(n) \leq 1.52165... \forall n$.  Use of
these bounds  allows us to delimitate the
critical parameters. Approximating $h(n)$ by one of the limiting values,
Eq.~(\ref{14}) can  be summed  explicitly,
$\sum_{n=2}^\infty n\rho_n = \rho-\rho_1$, reducing to a
single algebraic equation for $\rho_1$. Substituting the root of
this equation  back into 
Eq.~(\ref{14}) leads to the distribution of cluster sizes.
For large  $\lambda's$,  separation in two coexisting 
phases is once again found.  In fact, the critical densities and 
temperatures are identical to our earlier calculation done in the
absence of cluster formation. As $\lambda$ 
decreases, the critical temperature also diminishes, and dipolar association
becomes relevant.   We find
that below certain critical value, $ 0.07<\lambda_c<0.35$,
the phase separation 
disappears, see Fig. 1. Van Leeuwen and Smit 
could not locate phase separation for 
$\lambda < 0.3$, the value  which is very close to our upper bound.
We note, however, that the lower bound is probably more
realistic since, as can be seen from the inset to Fig 1,
the average chain size is still quite small in the region of
criticality.

We conclude that the theory presented above captures the essential
physics of criticality in polar fluids.
A number of issues still need to be addressed.  As it stands, the theory
does not take into account interactions between the chains. How these
residual forces can be introduced into the theory is far from
obvious.  We must stress, however, that the naive approach of allowing each
one of the monomers inside the chains  
to interact with  unrenormalized 
potential, Eq.~(\ref{1}), is incorrect\cite{Fis98}. 
This would double count the interactions, 
which are  partially taken care of by
the dipolar association, and would lead to phase separation for all values 
of $\lambda$.

The nature of criticality in polar fluids 
is also of great interest. Although there can be 
little doubt that the Stockmayer fluid ($\lambda=1$) belongs to the
Ising universality class, in view of our results, this no longer
seems to be so evident for dipolar fluids with $\lambda \approx 0.3$.
The fact that the mean-field theory can account so well for the
location of the critical point, suggests that for  $\lambda<0.5$ the
fluctuations play only a marginal role.  One might, therefore, expect
to see mean-field critical exponents, or a crossover from 
the mean-field to the Ising universality 
in a very narrow neighborhood of the critical
point.  As is the case for electrolytes, the criticality in dipolar
systems should remain a challenge for the foreseeable
future.

While this paper was going through the refereeing process,
new simulations\cite{Cam00} have once again raised a possibility of the
phase separation of $DHS$ at sufficiently low temperatures
and densities.  Unlike, the usual gas-liquid transition,
it has been argued, however, that this novel phase separation
is driven by the proliferation of Y-like bifurcations\cite{Tlu00}. 
The two coexisting phases are found to have respectively
a large and a small concentration of Y-like defects\cite{Pin00}.

This work was supported in part by  
Conselho Nacional de
Desenvolvimento Cient{\'\i}fico e Tecnol{\'o}gico (CNPq),  Financiadora 
de Estudos e Projetos (FINEP),
and by the National Science Foundation (NSF)
under Grant No. PHY94-07194.


\end{document}